\documentstyle{mn}
\input epsf

\newif\ifAMStwofonts

\title
{Limit shocks of relativistic magnetohydrodynamics, Punsly's 
waveguide and the Blandford-Znajek solution}
\author[S.S. Komissarov]
{
  S.S.Komissarov\\
Department of Applied Mathematics, 
The University of Leeds, 
Leeds LS2 9JT }

\begin{document}
\label{firstpage}
\maketitle

\begin{abstract}
In this paper we examine various issues closely related to the ongoing 
discussion on the nature of the Blandford-Znajek mechanism of extraction 
of rotational energy of black holes. In particular, 
we show that switch-on and switch-off shocks are allowed 
by the shock equations of relativistic MHD and have similar properties to 
their Newtonian counterparts. Just like in Newtonian MHD they are limits of  
fast and slow shock solutions and as such they may be classified as 
weakly evolutionary shocks. The analysis of Punsly's MHD waveguide problem 
shows that its solution cannot have the form of a traveling step 
wave and that both fast and Alfv\'en waves are essential for generating the 
flow in the guide. Causality considerations are used to argue that the 
Blandford-Znajek perturbative solution is in conflict with the membrane 
paradigm. An alternative interpretation is presented according to which     
the role of an effective unipolar inductor in the Blandford-Znajek mechanism is 
played by the ergospheric region of a rotating black hole. Various implications 
of this are discussed.   
   
\end{abstract}


\section{Introduction}

In coordinate systems singular at the black hole horizon, like the 
popular Boyer-Lindquist coordinates, the horizon is inevitably turns 
into a rather peculiar boundary of spacial domain. 
According to the widely accepted ``Membrane paradigm''  
the horizon, or rather somewhat less stringently defined ``stretched 
horizon'' which is placed somewhere just above the real horizon, may be 
identified with a rotating conducting sphere (e.g. Blandford 1979, 
Thorne et al. 1986). This makes magnetized black holes look analogous to 
magnetized neutron stars. For many years, beginning with \cite{PC}, 
Brian Punsly have been criticizing this view and also   
the perturbative steady-state electromagnetic wind solution 
for a force-free magnetosphere of a rotating black hole due to 
Blandford and Znajek \shortcite{BZ77}, the BZ solution, 
together with similar MHD models (e.g. Phinney 1982,1983) 
on the basis of causality arguments.  
Indeed, in the case of pulsars there is only  
an outgoing wind which passes first through the Alfv\'en critical surface 
and then through the fast critical surface and, thus, the neutron star can 
communicate with the wind by means of both fast and Alfv\'en waves 
(Since the gas pressure is dynamically insignificant 
in the tenuous magnetospheres of neutron stars and black holes, 
the slow waves seem to be irrelevant.)  Black holes, however, must also develop 
an ingoing wind which passes through its own pair of critical surfaces before 
reaching the horizon \cite{Jap90}. 
For the typical parameters of astrophysical black holes 
the inner fast surface is likely to be extremely close to the 
black hole horizon and one may argue that the stretched horizon can communicate 
with the outgoing wind by means of fast waves.  
On the other hand, the inner critical Alfv\'en surface may be rather  
distant from the black hole horizon. Thus, even the stretched horizon cannot 
communicate with the outgoing wind by means of Alfv\'en waves which makes 
black holes rather different from neutron stars. 

Blandford (1979) proposed that the outgoing wind of black holes is established 
by means of fast waves alone (This seems to be the only way to reconcile 
the membrane paradigm with the BZ solution.)  However, Punsly (1996,2001) argued 
that fast waves are completely irrelevant and Alfv\'en waves are solely 
responsible for creating the global system of poloidal 
electric currents of such winds and adjusting the angular velocity of 
magnetic field lines  and  suggested that the steady-state BZ solution is 
unstable and, hence, nonphysical. In particular, Punsly criticized  
Znajek's horizon boundary condition \cite{Z77} used to determine the wind 
constants in the BZ electrodynamic solution and in the MHD analysis of Phinney 
(1982,1983) as a condition imposed in a region causally disconnected from the 
outgoing wind. 

However, Znajek's condition is not really a boundary condition 
as it simply prohibits infinitely strong electromagnetic field as measured 
by a physical observer, e.g. a free falling observer (e.g. Phinney 1983).
Recently, Beskin and Kuznetsova \shortcite{BK00} stressed this point once 
more and argued that, though the interpretation of the stretched horizon 
as a unipolar inductor is misleading, there is nothing wrong with causality 
of the BZ-like MHD models.   This conclusion is strongly supported by 
the results of recent time-dependent electrodynamic simulations \cite{Kom01} 
which indicate asymptotic stability of the BZ solution (Znajek's boundary 
condition was not imposed in these simulations.) 

In this paper we  
continue the discussion of the causality paradox a bit further and attempt  
to clarify the nature of BZ mechanism. At first sight, the study presented 
here may appear rather unfocused but a closer look reveals strong 
connections between its sections. 
In Sec.2 we study limit shocks of relativistic MHD. 
In Sec.3 we use these results to analyse the Punsly's waveguide problem that 
provides important insights into the problem of relativistic MHD and 
electrodynamic winds.  
In Sec.4 we propose a modification of the membrane paradigm that 
does not conflict with causality.

\section{Limit shocks of relativistic MHD}

It is well known that shock equations of Newtonian MHD allow 
compressive solutions that have non-vanishing tangential component 
of magnetic field, ${\bmath B}_t$, only on one side of the discontinuity. These 
are called switch-on and switch-off shocks, e.g. \cite{JT64}. 
Both shocks propagate with Alfv\'en speed relative to the state 
with non-vanishing tangential component of the magnetic field. 
Relative to the other state, a switch-on shock is super-fast 
and a switch-off shock is sub-slow. It is also known that 
any switch-on solution  can be considered as a fast shock in 
the limit ${\bmath B}_t \to 0$ and any switch-off solution as a similar 
limit of a slow shock solution and this is why these shock are often 
called limit shocks \cite{JT64}. 

Strictly speaking,  switch-on shocks  
are not evolutionary simply because it is impossible to ensure that ${\bmath B}_t$ 
is exactly zero upstream of the shock. 
An infinitesimally small perturbation of the upstream state 
resulting in an infinitesimally small upstream tangential 
field will generally turn this shock into a fast shock followed by an Alfv\'en 
discontinuity. 
This fast shock will be infinitesimally close to the original switch-on shock in 
all respects except the direction of the downstream tangential component of magnetic 
field.  Moreover, both the fast and the Alfv\'en waves will have infinitesimally 
close wave speeds.  
Only a finite amplitude perturbation can result in a finite split of a 
switch-on shock. In this sense, switch-on shocks are similar to fast shocks    
but rather distinct from intermediate shocks of MHD which 
are genuinely non-evolutionary and split as a result of interaction with 
waves of infinitesimally small amplitude (e.g. Landau \& Lifshitz 1959, 
Falle and Komissarov 2001). Similar arguments apply to switch-off shocks. 
All these specific properties of limit shocks explain why they should 
be considered as physically meaningful solutions and why  
Jeffrey and Taniuti \shortcite{JT64} called them weakly evolutionary. 

Shock solutions of relativistic MHD have been a subject of rigorous 
analysis beginning with the pioneering paper by Hoffman and Teller 
\shortcite{HT50}. The results have been summarized in two rather 
comprehensive monographs by A.Lichnerowicz \shortcite{L67} and 
and A.M.Anile \shortcite{A89}. In  Lichnerowicz \shortcite{L67}
it was apparently shown that limit shocks do not exist in relativistic 
MHD. In Anile \shortcite{A89} and \cite{MA87} 
the analysis of limit shocks 
is not presented and readers are referred to the work by Lichnerowicz. 
It has to be stressed that, as it 
has been explained in Lichnerowicz \shortcite{L67}, 
the non-existence of limit shocks in 
relativistic MHD would make this system qualitatively different 
from Newtonian MHD. However, the author of this paper has recently 
found such shocks in numerical solutions of certain Riemann problems and 
realized that there must be a flaw in the analysis of Lichnerowicz. 

Since the style adopted in these monographs is a bit too mathematical 
we shall follow the more traditional approach of Hoffman and Teller \shortcite{HT50}.
First, we construct the limit shock solutions, then we show that 
there exist evolutionary shock solutions in the neighborhood of these 
solutions, and point out the error in the analysis of Lichnerowicz \shortcite{L67}. 
        
Let us define more precisely what is meant by 
the limit shocks in relativistic MHD. 
The first condition 
is that, in the rest frame of the fluid on one side of the shock  
the magnetic field is parallel to the shock normal.
This allows us to construct a shock frame  
that is moving with respect to the original fluid frame along the magnetic field.  
In this frame the shock is at rest, the electric field is vanishing and the 
fluid velocity is parallel to the magnetic field on both sides of the shock. 
This frame was first introduced 
by Hoffman and Teller \shortcite{HT50} to simplify the analysis of 
oblique MHD shocks. By construction, in this frame the magnetic field 
on one side of a limit shock is normal to the shock front.  
The second condition is a non-vanishing
tangential component of magnetic field on the other side. 
Thus, the shock can be described 
as either switching on or switching off the tangential component 
of magnetic field in the Hoffman-Teller frame.  

\begin{figure*} 
\leavevmode 
\epsffile[0 0 460 230]{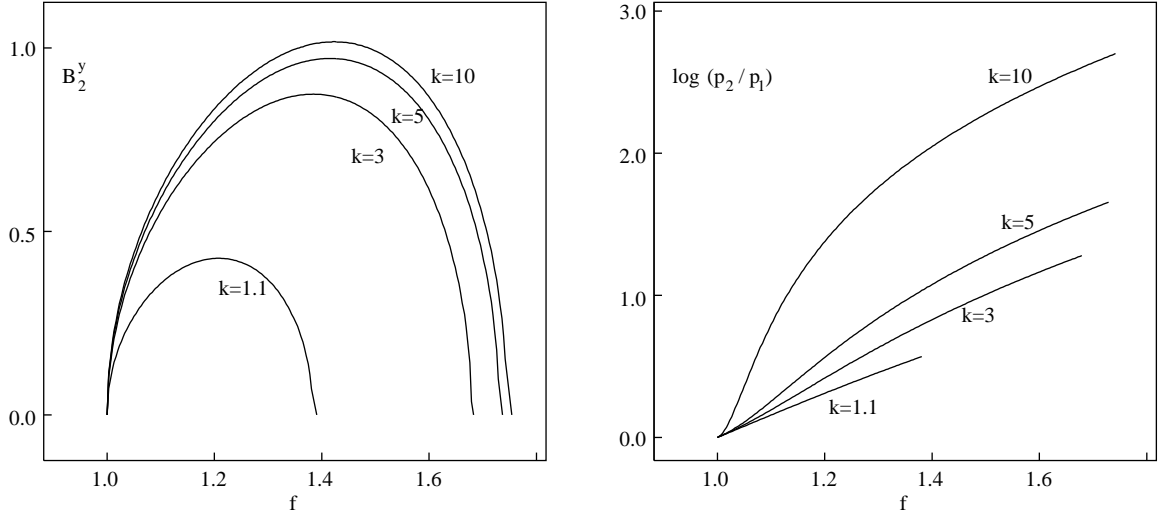} 
\caption{Switch-on shock solutions. Here $f=u^x_1/u_{a_1}$ and 
$k=u_{a_1}/u_{s_1}$. Index ``1'' refers to the state with vanishing $B^y$}. 
\label{fig_f}
\end{figure*}

\begin{figure*} 
\leavevmode 
\epsffile[0 0 460 230]{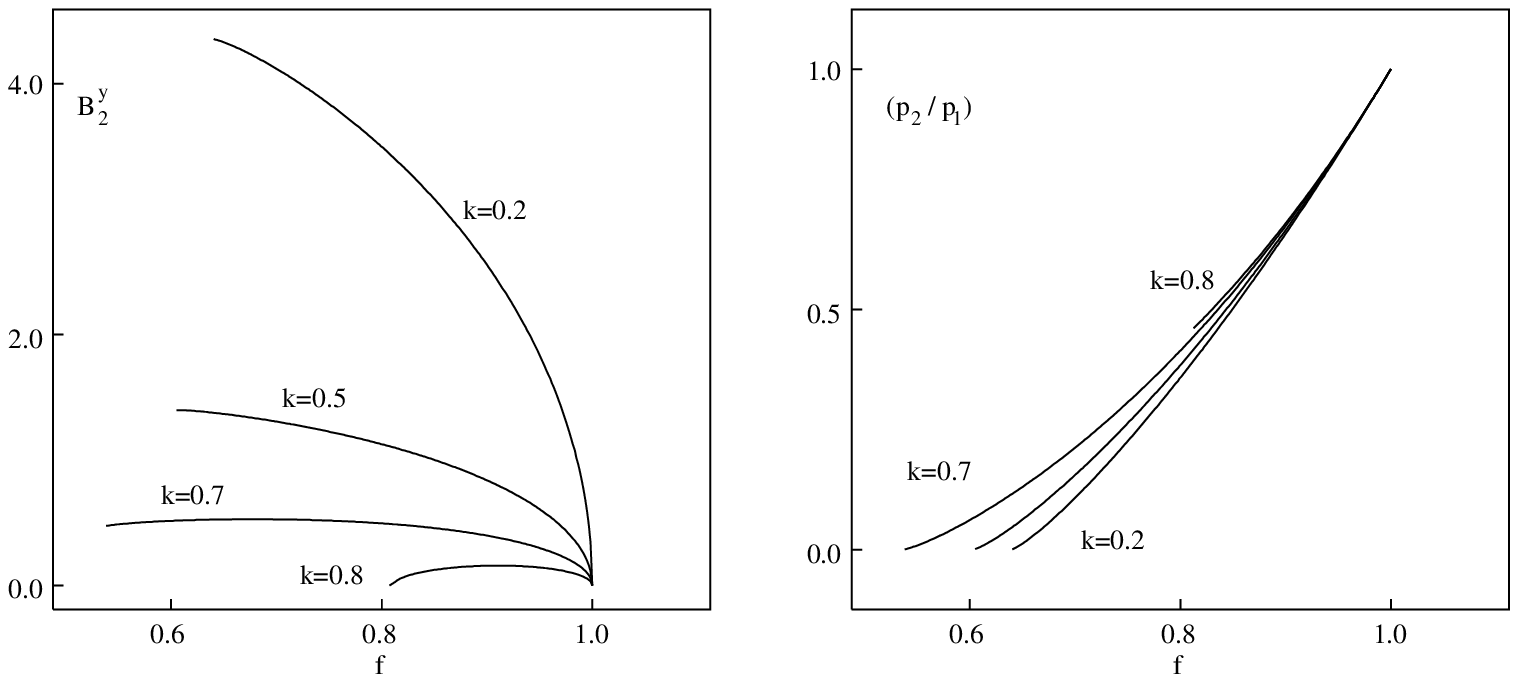} 
\caption{Switch-off shock solutions. Here $f=u^x_1/u_{a_1}$ and 
$k=u_{a_1}/u_{s_1}$. Index ``1'' refers to the state with vanishing $B^y$}. 
\label{fig_s}
\end{figure*}

\begin{figure*} 
\leavevmode 
\epsffile[0 0 460 460]{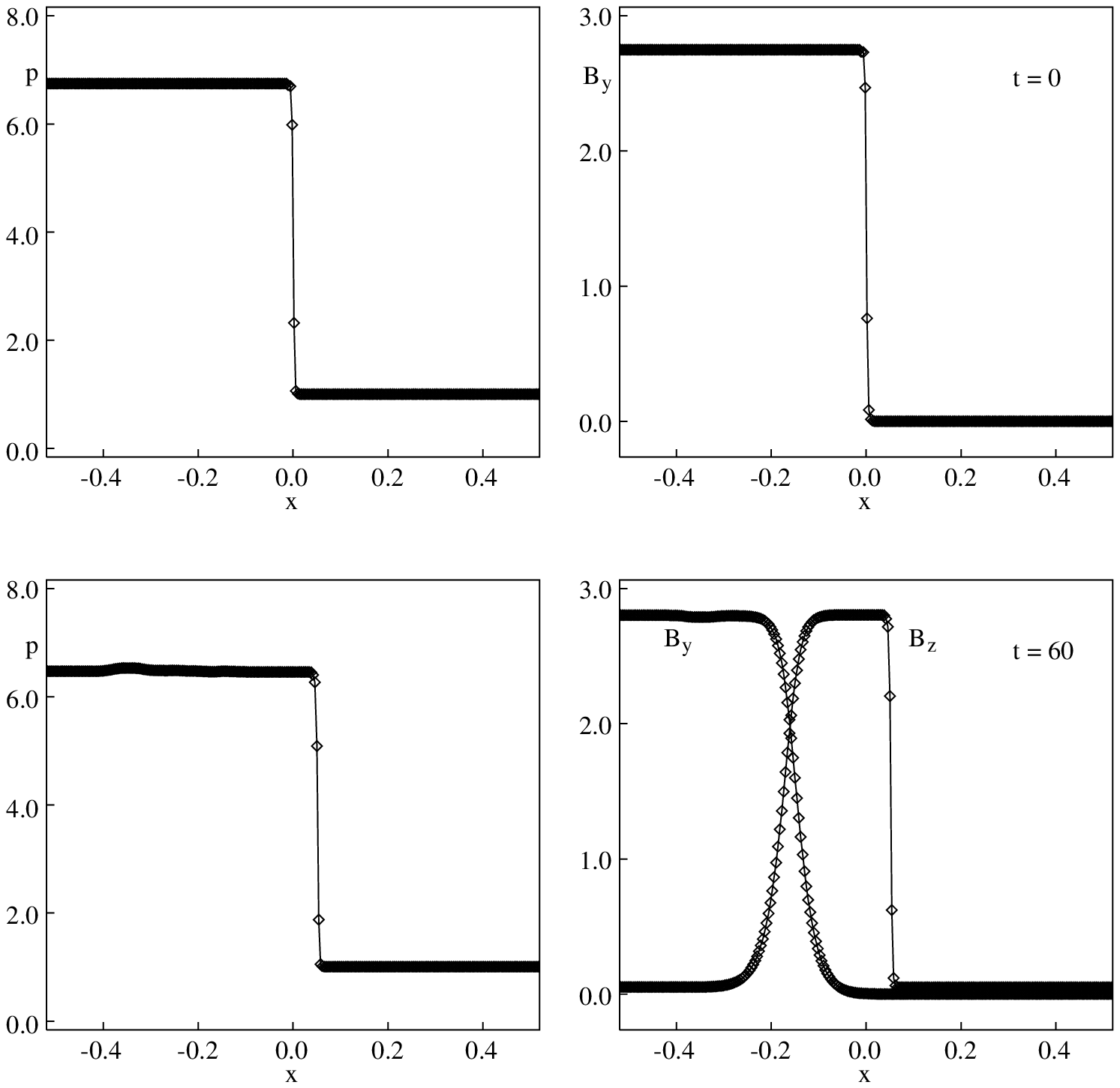} 
\caption{ Splitting of a switch-on shock into a fast shock and 
Alfv\'en wave as the result of interaction with a small perturbation 
of magnetic field introduced upstream of the initial solution. 
The top panels show the initial switch-on solution, where $p$ is the gas 
pressure. The shock is stationary relative to the computational grid. 
Its internal structure is due to artificial viscosity. The bottom 
panels show the result of interaction with a small perturbation of 
magnetic field ($B_z=0.05$ for $x>0.5$) upstream of the initial shock. 
The fast shock is located at $x \simeq 0.05$ and the Alfv\'en wave 
in $-0.3 < x < -0.1$.      
}
\label{fig_split}
\end{figure*}

From Maxwell equations it follows that in the Hoffman-Teller frame the 
electric field vanishes on both sides of the shock and, thus, the 
fluid velocity vector is parallel to the magnetic field vector 
\cite{HT50}: 
\begin{equation}
   u^i = s B^i, 
\label{uB}   
\end{equation}
where $u^i$ are the space components of the 4-velocity vector. 
Following \cite{HT50} we introduce Cartesian coordinates such that 
the x-axis is along the shock normal and $B^z=0$ on both sides of the shock.
Then the only shock equations we have to analyse
are 
\begin{description}
\item{Continuity equation:}
\begin{equation}
   [D]=0, 
\label{dD}   
\end{equation}
 
\item{Energy equation:}
\begin{equation}
   [T^{tx}]=0, 
\label{dT_tx}   
\end{equation}
       
\item{Momentum equations:}
\begin{equation}
[T^{xx}]=0, 
\label{dT_xx}   
\end{equation}
       
\begin{equation}
   [T^{xy}]=0, 
\label{dT_xy}   
\end{equation}
\end{description}
where for any quantity $A$ 
$$
   [A] =A_2-A_1,  
$$ 

\begin{equation}
   D=\rho u^x
\label{D}   
\end{equation} 

\begin{equation}
   T^{ti} = w u^t u^i,  
\label{T1}   
\end{equation}

\begin{equation}
   T^{ij} = w u^i u^j + (p+B^2/2) g^{ij} -B^iB^j.  
\label{T2}   
\end{equation}
Here $\rho$ is the rest mass density, $w$ is the relativistic enthalpy,
$p$ is the thermodynamic pressure, $u^\nu$ is the fluid 4-velocity, and we 
are using units such that $4\pi$, the magnetic permeability and the speed of 
light do not appear in the equations.    
From (\ref{dT_xy},\ref{T2},\ref{uB}) one has 

\begin{equation}
  [(ws^2-1)B^xB^y]=0. 
\label{key1}   
\end{equation}
Thus if $B^y=0$ only on one side of the shock then on the other side
\begin{equation}
  s^2=1/w. 
\label{s}   
\end{equation}
Combining (\ref{s}) with (\ref{uB}) one obtains 
\begin{equation} 
(u^t)^2=(w+B^2)/w,
\label{gamma}   
\end{equation}

\begin{equation} 
(u^x)^2 = (B^x)^2/w,
\label{ux}   
\end{equation}
and, finally, 
\begin{equation} 
(v^x)^2 = c_a^2 \equiv (B^x)^2/(w+B^2).  
\label{vx}   
\end{equation}

The last equation tells us that relative to the state with non-vanishing 
tangential field the shock propagates with Alfv\'en speed, $c_a$, 
e.g. \cite{Kom99}.
Substitution of these results into (\ref{dD}--\ref{dT_xx}) leads to  

\begin{equation} 
B^2 +2p = 2T^{xx},
\label{eq1}   
\end{equation}

\begin{equation}  
w-2p=\left(\frac{T^{tx}}{B^x}\right)^2 - 2T^{xx},
\label{eq2}   
\end{equation}
and
\begin{equation} 
\frac{\rho^2}{w} = \left(\frac{D}{B^x}\right)^2, 
\label{eq3}   
\end{equation}
where the shock invariants $D$, $T^{tx}$, and $T^{xx}$ may be evaluated
given the parameters of the state with vanishing $B^y$. 
Combined with the equation of state, 
(\ref{eq1},\ref{eq2}) allow us to determine the thermodynamic parameters on the 
other side of the shock. In the following we use index ``1'' for the state 
with vanishing $B^y$ and index ``2'' for the state with non-vanishing $B^y$. 

It is easy to verify that if $v_{x_1}=c_{a_1}$ then the shock vanishes.
Consider weak shocks with 
\begin{equation}
    u^2_{x_1}=u^2_{a_1}(1+\alpha), \quad |\alpha | \ll 1,
\label{pert}   
\end{equation}
where $u_a = c_a/\sqrt(1-c_a^2)$ may loosely be called the ``Alfv\'en 4-velocity''. 
Substituting this into (\ref{eq1}-\ref{eq3}) and retaining only terms 
linear in $\alpha$ one obtains  
\begin{equation} 
   B^2_{y_2}+2[p] = \alpha (2 w_1 u^2_{a_1}),
\label{weak1}   
\end{equation}

\begin{equation} 
   [w]-2[p] = \alpha w_1,
\label{weak2}   
\end{equation}

\begin{equation} 
   [\rho^2/w] = \alpha (\rho_1^2/w_1). 
\label{weak3}   
\end{equation}
From these and the second law of thermodynamics one has 

\begin{equation} 
   [p] = \alpha w_1 u^2_{s_1},  
\label{weak4}   
\end{equation}

\begin{equation} 
   B^2_{y_2} = 2 w_1 \alpha (u^2_{a_1}-u^2_{s_1}),  
\label{weak5}   
\end{equation}
where $u_s = a/\sqrt{1-a^2}$ is the ``sound 4-velocity'',   
$a$ is the sound speed. From (\ref{weak5}) one can see that 
weak limit shocks exist only if 
$$
 \alpha (c_{a_1}-a_1) >0.
$$ 
This can only be satisfied in the following two cases 

\begin{itemize} 
\item If $c_{a_1}>a_1$ and  $|v^x_1| > c_{a_1}$ then $p_2 > p_1$, 
this is a switch-on compressive shock,       
\item If $c_{a_1}<a_1$ and $|v^x_1| < c_{a_1}$ then $p_2 < p_1$, this is 
a switch-off compressive shock. 
\end{itemize}

Figure \ref{fig_f} shows the switch-on solutions for the polytropic 
equation of state with $\gamma =4/3$, 
$\rho_1=1$, $p_1=1$ and $k=u_{a_1}/u_{s_1}=1.1,2,3,10$.  
Like in Newtonian MHD \cite{JT64}, a relativistic switch-on shock turns into 
a pure gas dynamical shock propagating along the magnetic field if the shock 
speed exceeds a certain critical value. 
    
Figure \ref{fig_s} shows the switch-off solutions 
for the polytropic equation of state 
with $\gamma =4/3$, $\rho_1=1$, $p_1=1$ and $k=u_{a_1}/u_{s_1}=0.8,0.7,0.5,0.2$. 
For $k=0.8$ the switch-off shocks turns into a pure gas dynamical shock. 
For other values of $k$ the shock curve terminates as the gas pressure 
of the upstream state vanishes. 

The evolutionary conditions for shock solutions are of very general nature 
and apply to relativistic shocks in exactly the same manner as to Newtonian 
shocks. Thus, evolutionary compressive relativistic MHD shocks must be either 
super- or sub-Alfv\'enic on both sides of the shock \cite{FK01}. 
Let us show that in the neighborhood of limit shocks there exist 
evolutionary shock solutions.  From (\ref{key1}) one can see that if 
$B^y$ has the same sign on both sides of the shock then so does

\begin{equation}  
\mu = ws^2-1.  
\label{mu}   
\end{equation}
From (\ref{uB}) one obtains 
\begin{equation}  
v_x^2 = f(\mu)=\frac{(1+\mu)B_x^2}{w+(1+\mu)B^2}. 
\label{vx2}   
\end{equation}
Since $f'(\mu)>0$, one immediately concludes that 
\begin{itemize}
\item If $\mu>0$ then $v_x^2 > c_x^2$ on both sides of the shock,
\item If $\mu<0$ then $v_x^2 < c_x^2$ on both sides of the shock,
\end{itemize} 
and, thus, the shock is evolutionary. For a switch-on shock
$\mu_2>0$. Thus, if we introduce an infinitesimally small $B^y_1$ of the 
same sign as $B^y_2$ the new shock solution will be a fast shock, 
just like in Newtonian MHD \cite{JT64}. 
Similarly, a switch-off shock turns into a slow shock.    

Due to the general nature of evolutionary conditions splitting of not 
strictly evolutionary shocks must proceed in the same fashion regardless 
of relativistic or classical nature of governing equations. 
Figure~\ref{fig_split} shows the effect of a small perturbation of the 
upstream magnetic field on a switch-on shock. The initial solution   

\begin{displaymath}
\begin{array}{ll}
   \mbox{Left State:} \quad & \mbox{Right state:} \\
   \bmath{B}=(3.149,2.749,0), \quad & \bmath{B}=(3.149,0,0), \\
   \bmath{E}=0, \quad &  \bmath{E}=0,  \\
    \rho=3.454, \quad p=6.743, \quad & \rho=1, \quad p=1 .
\end{array}
\end{displaymath}

describes a stationary switch-on shock. It internal structure seen in 
fig.~\ref{fig_split} is entirely due to artificial viscosity. The upstream
state is then perturbed by introducing tangential magnetic field with 
$B_z=0.05$ upstream of the shock. The eventual interaction splits the 
switch-on shock into a fast shock and Alfv\'en wave in the same manner as 
it has been described in Sec.1.  In these simulation we used polytropic 
equation of state with the ratio of specific heats $\gamma=4/3$. 
These simulation were carried out using the upwind numerical scheme 
described in \cite{Kom99}.  

\begin{figure*} 
\leavevmode 
\epsffile[0 0 300 200]{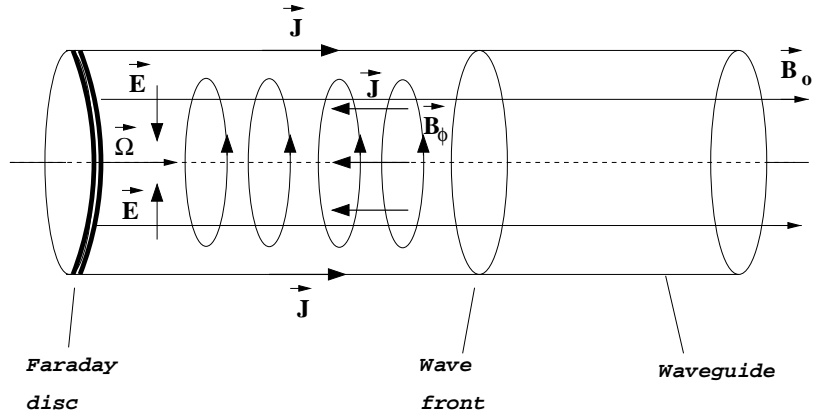} 
\caption{The electromagnetic properties of the flow pattern in a cylindrical 
waveguide attached to a rotating conducting disc as described in Punsly (2001).}
\label{fig-cyl}
\end{figure*}

Finally, we find necessary to explain the error in the analysis 
of Lichnerowicz \shortcite{L67} that eventually led to the incorrect 
conclusion on non-existence of the limits shock solutions (p.161). 
On page 151 of the book, equation (47-1), a space-like 4-vector $h^\alpha$ is 
decomposed into a sum of two 4-vectors, one parallel and the other one 
normal to a unit space-like vector $n^\alpha$:  
$$
    h^\alpha=t^\alpha - \eta n^\alpha, \quad t^\nu n_\nu =0. 
$$   
Then it is claimed that $t^\alpha$ is always space-like. However, this is 
not true. The reader can easily verify that if   
$h^\alpha = (0,h,0,0) $ and  $n^\alpha=(\sqrt{3},2,0,0)$ then $t^\alpha$ is 
time-like.

\section{Punsly's waveguide problem}    

In order to demonstrate the exceptional role of Alfv\'en waves, 
Punsly proposed to consider a much simpler problem 
involving a Faraday wheel connected to a cylindrical waveguide uniformly 
filled with cold  tenuous plasma and a strong axial magnetic field 
(Sec.2.9.4 in Punsly 2001). 
For this problem he claimed to have constructed a step Alfv\'en wave 
solution of MHD equations corresponding to an instantaneous  spinning up of 
the disc. Downstream of the wave front the magnetic field has only axial 
and azimuthal components, the electric field is radial and the electric 
current is axial (see figure~\ref{fig-cyl}). The return current flows over 
the surface of the guide and the global current closure is ensured 
by the displacement current of the leading front. Such simple step wave solution, 
however, is impossible for the following simple reason.  

Since in the unperturbed state the 
Alfv\'en speed in the direction of magnetic field is uniform across the waveguide, 
an Alfv\'en front launched from the surface of the disc would indeed stay normal 
to the guide axis. The problem is that such wave would have no effect on the 
state of plasma,  one could call it a ghost wave. 
Indeed, a frame propagating with Alfv\'en speed in the axial direction 
is the Hoffmann-Teller frame, as in this frame ${\bmath E}=0$ upstream. 
However, in the Hoffmann-Teller frame the amplitude of 
the tangential component of magnetic field remains unchanged by an Alfv\'en 
shock \cite{Kom97} just like it does in Newtonian MHD. 
Since in this case the tangential component is zero upstream it must be vanishing 
downstream as well.  It is easy to verify that the Lorentz 
transformation to the original frame of the waveguide preserves this result. 
Thus, the Alfv\'en shock has to be ruled out as well as the claim that 
the solution to this problem involves only Alfv\'en waves. 

The tangential component of magnetic can be generated by a switch-on fast 
compression wave including a switch-on shock. 
However, a switch-on shock cannot not remain plane given the properties 
of the ``solution'' downstream of the discontinuity. Indeed, for a constant 
shock speed relative to the unperturbed state the tangential component of the 
downstream magnetic field would have the same amplitude everywhere 
(see eqs \ref{eq1}-\ref{eq3}) whereas in the guide it must vanish
along the symmetry axis.   
Thus, the solution to the Punsly's problem in the framework of 
relativistic MHD cannot have the simple form of a step wave altogether. 

Although we have not been able to find an analytical solution to the waveguide 
problem, the following basic arguments suggest that {\it both fast and Alfv\'en 
waves are important in this problem}. 
When the disc rotation is switched on, the induced velocity shear generates 
the azimuthal magnetic field $ B_\phi \propto r$ at the disc surface. Since the 
force balance is broken a switch-on MHD shock is driven into the guide.   
Initially this shock is plane but since the shock speed depends on $r$ 
it eventually becomes curved.  In the local Hoffman-Teller frame of such 
curved discontinuity the upstream tangential 
component of magnetic field will generally have a direction which is different 
from the one of the downstream tangential component and, thus, 
the discontinuity will no longer satisfy the MHD shock equations. 
As explained in Sec.2 it will split mainly into 
a fast shock followed by an Alfv\'en wave.  
{\it Thus, the disc ultimately emits both fast and Alfv\'en waves.} 
Exactly the same evolution would be observed in the case of Newtonian MHD.  
The role of fast waves is to create and amplify the non-axial field whereas 
the Alfv\'en waves ensure that it becomes azimuthal. 
   
In the limit of force-free degenerate electrodynamics \cite{Kom02,B02} 
the wavespeed of the fast wave tends to the speed of light and 
this makes a step wave solution possible. Indeed, it is easy to verify that 
equations of degenerate electrodynamics allow the following traveling wave 
solution: 
\begin{eqnarray} 
\nonumber
 & & B_r=0,\quad B_\phi(t,z)=-\Omega (t-z) r B_0,\quad B_z=B_0, \\
 & & E_r(t,z)=B_\phi(t,z),\quad E_\phi=0,\quad E_z=0, 
\end{eqnarray}
where the components of vectors are given in the orthonormal basis of 
cylindrical coordinates and $\Omega(t)$ is the angular velocity of the disc. 
One could be tempted to interpret the discontinuity of such step-wave solution 
as a limit of a switch-on shock and call it a fast wave. However,      
the eigensystem of degenerate electrodynamics degenerates 
in the axial direction (in general, in the  direction of 
$\bmath{E}\times\bmath{B} \pm \bmath{B}\sqrt{B^2-E^2}$ 
\cite{Kom02}), that is both the Alfv\'en  and the fast linear waves propagate 
with the same speed, which is the speed of light, and are no longer distinct. 
Moreover, both modes are linearly degenerate and, thus, their shock solutions 
have the same properties as the small amplitude waves. This makes it quite  
impossible to tell whether this solution represents a fast or 
an Alfv\'en wave of degenerate electrodynamics. In fact, it should  
be regarded as a mixture of both modes. Indeed, in the waveguide which is not 
perfectly cylindrical this degeneracy will be removed and the solution will 
split into two distinct waves in the way similar to splitting of a switch-on 
shock discussed above.

\section{Membrane paradigm and the Blandford-Znajek solution}

The results of the previous section show that in Punsly's MHD waveguide 
problem both fast and Alfven waves are important. 
This has to be true in general including the case  
of black holes magnetospheres. In such magnetospheres Alfv\'en waves propagate 
only along the poloidal field lines (Appendix A). 
Thus, fast waves must play a special role in establishing the cross-field balance 
whereas Alfv\'en waves are particularly important for establishing the 
wind constants (They are constant along the poloidal field lines.)       

Since neither of the waves can be ignored  
the causality paradox posed by Punsly and Coroniti \shortcite{PC} has to be
taken seriously. The resolution of the paradox proposed by Punsly and Coroniti  
involves rejection of both 1) the membrane paradigm or rather its part that 
identifies the stretched horizon with the rotating conducting surface 
of a unipolar inductor and 2) the Blandford-Znajek solution.  
They eventually tried to construct completely different models of black hole 
magnetospheres involving rotating dense shells and discs of accreting matter 
as material analogues of the Faraday disc \cite{P01}.  
However there seems to exist a different resolution of this paradox which we 
discuss here.   

In fact, the causality 
arguments directly hit the membrane paradigm and it cannot be saved. 
The analogy between the horizon and a rotating conducting sphere is not at 
all that complete as it is believed. It does allow a simplified presentation 
for an audience unfamiliar with general relativity but it does not really 
provide deep insights into the physics of black hole magnetospheres.  

Things are different when it comes to the BZ solution. 
{\it All what the causality 
paradox tells us is that this solution is inconsistent with the membrane 
paradigm!} 
Historically, the BZ solution was used to build the paradigm and now it 
is widely considered that both are inseparable, but in fact 
the BZ solution clearly indicates the limitations of the paradigm. 
Clearly, there is no 
material analog of the Faraday disc in the BZ magnetosphere. 
If the horizon does not play its role then what is forcing the rotation 
of magnetic field lines?  
The answer can only lay in the properties of space-time outside of the black 
hole horizon as it has been suggested in \cite{B02}. Since another energy 
extraction mechanism, namely the Penrose mechanism \cite{P69},  
operates only within the ergosphere it seems reasonable to consider the possibility 
that {\it it is the ergospheric region of space-time that serves as a driving 
``force'' of the BZ mechanism.}  In fact, this is consistent with the causality 
arguments.  
  
The super-Alfv\'enic region of the ingoing wind 
cannot and does not have to communicate with 
the outgoing wind by means of Alfv\'en waves just like the super-Alfv\'enic 
region of the outgoing wind cannot communicate with the ingoing wind in such a 
way.   
However, the driving source responsible for both the outgoing and the ingoing winds
must be able to communicate with the winds by means of both fast and Alfv\'en waves 
and, thus, it must be located between the Alfv\'en surfaces and, thus, well outside 
of the horizon (Similar conclusion was reached by Punsly \shortcite{P01} and 
Beskin \& Kuznetsova \shortcite{BK00}.) 
The position of these critical surfaces, which merge with the light surfaces 
in the limit of degenerate electrodynamics (Appendix A), 
is not fixed as it depends on the angular velocity of magnetic
field lines, $\Omega_f$, which depends on many factors including the interaction 
with the surrounding plasma (effective load of the black hole electric circuit). 
However, for all values of $\Omega_f$ 
consistent with extraction of energy of a black hole in the BZ solution 
the inner Alfv\'en surface is located inside 
the ergosphere (Appendix A). The only exception is the polar direction 
where the light surface, the horizon, and the ergosphere coincide, but 
the Poynting flux density vanishes along the symmetry axis and, 
thus, there is no outgoing wind as well. {\it Thus, there always exists an 
outer region of the ergosphere which is    
causally connected to the outgoing wind of the BZ solution.} 
    
   In order to verify this conjecture one could study the dynamical behaviour 
of magnetic field lines that do not penetrate 
the horizon. The lines that enter the ergosphere are expected to be forced into 
rotation whereas those that do not should remain nonrotating.   
Such study is under way.   

Naturally, the driving force has to be the same both in electrodynamic and 
MHD models. In MHD approximation the wavespeeds of both fast 
and Alfv\'en waves are smaller then the speed of light and therefore the 
corresponding inner critical surfaces lay strictly outside of the horizon 
everywhere. Since the horizon and the ergosphere always coincide in the 
polar direction, both inner critical surfaces are situated outside of the 
ergosphere in the polar region. 
Thus, there exists a polar flux tube which is causally disconnected from the 
ergosphere. Within such a tube there can only be possible an accretion. 
Such conclusion is not entirely unexpected as a particle located on the symmetry 
axis is subject only to gravitational attraction.         

Another related issue is whether the approximation of degenerate electrodynamics 
(or magnetically dominated ideal MHD) brakes down somewhere near the horizon. 
It is impossible to get full answer to this question within the framework of 
degenerate electrodynamics. For example, if the condition 
$\bmath{E} \cdot \bmath{B}=0$ is 
satisfied by the initial solution it will be preserved during the evolution though 
there may not be enough charged particles to ensure this condition.   
However, the preservation of the other condition $B^2-E^2>0$, which is required 
for the hyperbolicity of degenerate electrodynamics, is not guaranteed \cite{Kom02}. 
As $B^2\rightarrow E^2 \not=0$ the drift velocity of plasma tends to the speed of 
light indicating  that particle inertia may need to be taken into account. However,  
the BZ solution satisfies this condition all the way to the horizon and 
the inertial effects are unlikely to be important along the magnetic field 
lines threading the horizon. This is not so obvious in the case of magnetic field 
lines threading the equatorial plane of the ergosphere where the approximation of
degenerate electrodynamics or ideal MHD has to break down in order to ensure the 
current closure condition. Punsly and Coroniti \shortcite{PC2} argue that particle 
inertia becomes important in the equatorial region as plasma is forced into 
rotation with almost speed of light relative to the zero angular velocity 
observers. However, they have not taken into account the Compton drag which 
may well lead to a much lower Lorentz factor of plasma rotation. This problem 
requires further investigation. 

\section{Conclusions}

\begin{enumerate}
\item
Switch-on and switch-off shocks are allowed 
by the shock equations of relativistic MHD and have similar properties to 
their Newtonian counterparts. Just like in Newtonian MHD they are limits  
of fast and slow shock solutions and as such they may be classified as 
weakly evolutionary shocks.  

\item
Contrary to what is claimed in \cite{P01},
the solution to Punsly's MHD waveguide problem cannot have the form of a 
step-like traveling wave and the guide flow cannot be established 
by means of Alfv\'en waves alone.  Even in the limit of degenerate 
electrodynamics where a step-wave solution exists it involves a mixture both 
fast and Alfv\'en waves. This suggests that both waves are important in 
the problems of magnetically driven MHD and electrodynamic winds. 

\item  
Blandford-Znajek solution contradicts to the membrane paradigm as the 
stretched horizon cannot play the role of a unipolar inductor. Causality 
arguments suggest that, just like in the case of the Penrose mechanism,    
the driving ``force'' of the Blandford-Znajek mechanism is the ergospheric 
region of space-time.         
\end{enumerate}

\section*{Acknowledgements}

The author thanks Brian Punsly for useful, though sometimes heated, 
discussions of the waveguide problem and various aspects of the   
magnetohydrodynamics of black holes and Sam Falle for his valuable  
comments on these and related issues.

\appendix
\section{} 

The wavespeed of Alfv\'en waves of degenerate electrodynamics is given by  

\begin{equation} 
   \mu_{\pm} = \bmath{\mu}_\pm \cdot \bmath{n}
\label{a1}
\end{equation} 

\noindent
where $\bmath{n}$ is a unit vector normal to the wave front and 
 
\begin{equation} 
    \bmath{\mu}_\pm = \frac{1}{B^2} \left(
     \bmath{E}\times\bmath{B} \pm \bmath{B}\sqrt{B^2-E^2} 
     \right) 
\label{a2}
\end{equation} 

\noindent
\cite{Kom02}. On a critical surface with normal $\bmath{n}$  
either $\mu_+$ or $\mu_-$ given by \ref{a1} vanishes

\begin{equation} 
   \bmath{\mu}_\pm \cdot \bmath{n} =0 
\label{a2b}
\end{equation} 

\noindent
Consider the Kerr metric in the Boyer-Lindquist coordinates 
$\{t,\phi,r,\theta\}$. For a steady-state force-free axisymmetric magnetosphere 
$\bmath{n}$ has only poloidal component and 
\begin{equation} 
    F_{tk} =-\Omega_f F_{\phi k}, \quad F_{t\phi}=0, 
\label{a3} 
\end{equation} 

\noindent
where $\bmath{F}$ is the electromagnetic field tensor, $\Omega_f$ 
is the angular velocity of magnetic field lines, and $k=r,\theta$ \cite{BZ77}. 
Then in the orthonormal basis 
$\{\bmath{e}_{\hat\phi}, \bmath{e}_{\hat r}, \bmath{e}_{\hat\theta} \}$ 
of a local fiducial observer, FIDO, one has

\begin{equation} 
E_{\hat\phi}=0,\quad E_{\hat r}=K B_{\hat\phi}, 
\quad E_{\hat\theta}=-K B_{\hat r}, 
\label{a4} 
\end{equation} 

\noindent
where 

\begin{equation} 
  K = \frac{\sqrt{g_{\phi\phi}}}{\alpha} (\Omega_f-\Omega_F),  
\label{a5} 
\end{equation} 

\noindent
$\alpha$ is the ``lapse function'', and 
$\Omega_F =-g_{\phi t}/g_{\phi\phi}$ is the FIDO's angular velocity.  
From \ref{a2},\ref{a4} one obtains the poloidal component of 
$\bmath{\mu}_\pm$:  

\begin{equation} 
    \bmath{\mu}_{p_\pm} = \frac{\bmath{B}_p}{B^2} \left(
   -K B_{\hat\phi} \pm \sqrt{B^2-K^2B_p^2} \right), 
\label{a6} 
\end{equation} 

\noindent
where $\bmath{B}_p$ is the poloidal magnetic field.
Thus, the condition \ref{a2b} is satisfied if either  

\begin{equation} 
   (\bmath{B}_p\cdot\bmath{n}) = 0, 
\label{a7} 
\end{equation} 

or

\begin{equation} 
   \bmath{\mu}_{p_\pm} =0.  
\label{a8} 
\end{equation} 

\noindent
\ref{a7} simply states that Alfv\'en waves propagate only 
along the poloidal magnetic field lines whereas \ref{a8} is the 
sought criticality  condition. For $\bmath{B}_p \not=0$ this condition 
can be written in terms of $\Omega_f$ and components of the metric tensor 
as 

\begin{equation} 
 f(\Omega_f,r,\theta)= g_{\phi\phi}\Omega_f^2 +2g_{t\phi}\Omega_f + g_{tt} = 0, 
\label{a9} 
\end{equation} 

\noindent
which is the well known equation of a light surface \cite{Jap90}. 
  
Without any loss of generality we may assume that $\bmath{B}_p$ is outgoing 
and consider only the northern hemisphere.  
In this case the condition 

\begin{equation} 
   \bmath{\mu}_{p_+} =0  
\label{a10} 
\end{equation} 
corresponds to the inner critical surface whereas

\begin{equation} 
   \bmath{\mu}_{p_-} =0  
\label{a11} 
\end{equation} 

\noindent
corresponds to the outer critical surface. From \cite{BZ77} we find that 
in the case of outgoing energy flow 
\begin{equation} 
   \frac{d \cal E}{d\Phi} = -B_T \Omega_f >0 ,
\label{a12} 
\end{equation} 

\noindent
where $\cal E$ is the energy flux within a flux tube of magnetic flux 
$\Phi$ and 
\begin{equation} 
   B_T = \sqrt{-g}F^{r\theta} = \alpha \sqrt{g_{\phi\phi}} B_{\hat\phi}. 
\label{a13} 
\end{equation} 

\noindent
Thus, the energy is extracted from the black hole only if 
\begin{equation} 
   B_{\hat\phi}\Omega_f <0.   
\label{a14} 
\end{equation}  

\noindent
Given this condition, \ref{a6} and \ref{a10} show that at the inner
critical surface 
\begin{equation} 
   \Omega_f (\Omega_f -\Omega_F) < 0.    
\label{a15} 
\end{equation}  

\noindent
For a black hole with positive angular velocity this means  
\begin{equation} 
  0 < \Omega_f < \Omega_F.     
\label{a16} 
\end{equation}  

\noindent
Thus, at the inner critical surface the magnetic field lines rotate 
slower than local FIDOs. Similarly one shows that at the outer 
critical surface the field lines rotate faster then local FIDOs
\begin{equation} 
  0 < \Omega_F < \Omega_f.      
\label{a17} 
\end{equation}  

On the surface of the ergosphere 

\begin{equation} 
 f(\Omega_f,r,\theta)= g_{\phi\phi}\Omega_f (\Omega_f -2\Omega_F).
\label{a18} 
\end{equation} 

\noindent
From this one can see that in the limit $\Omega_f \rightarrow 0$ the 
inner light surface coincides with the ergosphere, $f$ being positive 
inside and negative outside. Since

\begin{equation} 
 \frac{\partial f}{\partial \Omega_f}= 2 g_{\phi\phi}(\Omega_f -\Omega_F) 
  < 0 \quad\mbox{for } \Omega_f=0, \, \theta \not=0
\label{a19} 
\end{equation} 

\noindent
the inner critical surface moves inside the ergosphere as $\Omega_f$ 
increases and must remain inside for all values satisfying \ref{a16} 
(The third factor in \ref{a18} vanishes only when the outer critical 
surface moves inside the ergosphere.)


\begin{thebibliography}{} 

\bibitem[\protect\citename{Anile }1989]{A89}
Anile A.M., 1989, ``Relativistic Fluids and Magnetofluids'', Cambridge Univ.
Press, Cambridge.

\bibitem[\protect\citename{Beskin \& Kuznetsova }2000]{BK00} 
 Beskin V.S. and Kusnetsova I.V., 2000, Nuovo Cimento B, 115, 795.

\bibitem[\protect\citename{Blandford }1979]{B79} 
 Blandford R.D. 1979, in ``Active Galactic Nuclei'', ed. C.Hazard and 
 S.Mitton, Cambridge Univ. Press, Cambridge. 

\bibitem[\protect\citename{Blandford }2002]{B02} 
 Blandford R.D. 2002, astro-ph/0202265. 

\bibitem[\protect\citename{Blandford \& Znajek }1977]{BZ77} 
 Blandford R.D. and R.L. Znajek R.L., 1977, MNRAS, 179, 433  

\bibitem[\protect\citename{Falle and Komissarov }2001]{FK01}
Falle S.A.E.G. and Komissarov S.S., 2001, J.Plasma Phys., 65, 29. 

\bibitem[\protect\citename{Hoffmann \& Teller }1950]{HT50}
de Hoffmann F. and Teller E., 1950, Phys. Rev., 80, 692 

\bibitem[\protect\citename{Jeffrey and Taniuti }1964]{JT64}
Jeffrey A. and Taniuti T., 1964, ``Non-linear wave propagation'', 
Academic Press, New York London. 

\bibitem[\protect\citename{Komissarov }1997]{Kom97} 
 Komissarov S.S., 1997, Phys.Lett.A, 232, 435.   

\bibitem[\protect\citename{Komissarov }1999]{Kom99} 
 Komissarov S.S., 1999, MNRAS, 303, 343  

\bibitem[\protect\citename{Komissarov }2001]{Kom01} 
 Komissarov S.S., 2001, MNRAS, 326, L41.  

\bibitem[\protect\citename{Komissarov }2002]{Kom02} 
 Komissarov S.S., 2002, MNRAS, in press (astro-ph/0202447).  

\bibitem[\protect\citename{Landau \& Lifshitz }1959]{LL59} 
 Landau L.D. and Lifshitz E.M., 1959, ``Fluid Mechanics'', 
 Pergamon Press, Oxford.  

\bibitem[\protect\citename{Lichnerowicz }1967]{L67}
Lichnerowicz A., 1967, ``Relativistic Hydrodynamics and Magnetohydrodynamics'',
, Benjamin, New York.   

\bibitem[\protect\citename{Majorana and Anile }1987]{MA87}
Majorana A. and Anile A.M., 1987, Phys.Fluids, 30, 3045.

\bibitem[\protect\citename{Penrose }1969]{P69} 
 Penrose R., 1969, Rev.Nuovo Cim., 1, 252. 

\bibitem[\protect\citename{Phinney }1982]{Ph82} 
 Phinney E.S., 1982, in Ferrari A. and Pacholczyk A.G. eds, 
 Astrophysical Jets, Reidel, Dordrecht 

\bibitem[\protect\citename{Phinney }1983]{Ph83} 
 Phinney E.S., 1983, Ph.D. thesis, Univ.Cambridge. 

\bibitem[\protect\citename{Punsly }2001]{P01}
Punsly B., 2001, ``Black Hole Gravitohydromagnetics'', Springer-Verlag,
Berlin.  

\bibitem[\protect\citename{Punsly }1996]{P96}
Punsly B., 1996, ApJ., 467, 105.
 
\bibitem[\protect\citename{Punsly \& Coroniti }1990a]{PC} 
 Punsly B. and Coroniti F.V., 1990a, ApJ., 350, 518. 

\bibitem[\protect\citename{Punsly \& Coroniti }1990b]{PC2} 
 Punsly B. and Coroniti F.V., 1990b, ApJ., 354, 583. 

\bibitem[\protect\citename{Takahashi et al. }1990]{Jap90} 
 Takahashi M., Nitta S., Tatematsu Y., and Tomimatsu A.,
 1990, ApJ., 363, 206. 

\bibitem[\protect\citename{Thorne et al. }1986]{Membr} 
 Thorpe K.S., Price R.H., and Macdonald D.A., 1986, 
``The Membrane Paradigm'', Yale Univ. Press, New Haven.   

\bibitem[\protect\citename{Znajek }1977]{Z77} 
 Znajek R.L., 1977, MNRAS, 179, 457.   

\end{thebibliography}
\end{document}